# Modeling mammalian gastrulation with embryonic stem cells

Eric D. Siggia[1], Aryeh Warmflash[2]

[1]Center for Studies in Physics and Biology, The Rockefeller University, New York, NY 10065
[2]Departments of Biosciences and Bioengineering, Rice University, Houston, TX 77005

**Abstract**
Understanding cell fate patterning and morphogenesis in the mammalian embryo remains a formidable challenge. Recently, in vivo models based on embryonic stem cells (ESCs) have emerged as complementary methods to quantitatively dissect the physical and molecular processes that shape the embryo. Here we review recent developments in using embryonic stem cells to create both two and three-dimensional culture models that shed light on mammalian gastrulation.

## Introduction

During embryogenesis, the processes of cell differentiation, growth, division, and movement all occur simultaneously in a three-dimensional environment. For the mammalian embryo, this occurs in utero. The complexity of studying this process makes it crucial to develop simplified systems where these processes can be separated, readily observed, and studied in a controlled manner. Further, when considering human embryogenesis, the nearly completely lack of access to actual embryos means that developing synthetic systems may be the best route to understanding uniquely human features of early development. In this review, we focus on modeling mammalian development at gastrulation stages with systems derived from embryonic stem cells. We will consider both two and three-dimensional culture systems and will focus on the most recent developments. The reader is referred to other reviews for further discussion of earlier work [1-4].

Gastrulation occurs in the posterior region of the embryo under the control of signals emanating from two extraembryonic tissues, the visceral endoderm and the trophectoderm [5]. BMP signals from the trophectoderm initiate gastrulation at the proximal end of the embryo, while inhibitors to BMP, Nodal and Wnt are secreted from the anterior visceral endoderm and ensure that the site of gastrulation, known as the primitive streak, is confined to the posterior side of the embryo [5] [6]. The BMP signals are triggered by Nodal signals initiating from the epiblast, and in turn activate Wnt signals in the epiblast which further activate Nodal [7]. These high levels of Wnt and Nodal in the primitive streak are essential for gastrulation, and Nodal is thought to pattern the resulting mesendoderm in a dose-dependent manner with the highest levels of Nodal being required for endoderm and axial mesoderm and lower levels giving rise to paraxial and lateral mesoderm [8]. These facts have largely been inferred from the patterns of expression and knockout phenotypes of pathway components, and understanding the relationship between BMP, Wnt and Nodal signals and the resulting cell fates remains a challenge. Further, how the potential gradients of Wnt and Nodal activity are established and interpreted remains largely obscure.

ESCs offer an exciting window into mammalian development and have been used to model a wide variety of cell fate decisions and differentiation programs that take place in early embryogensis (e.g. [9-12]). Until recently, there were no methods to generate reproducible patterns from ESCs, and most protocols have either been highly optimized to produce a single cell type (e.g. [9,12,13]) or else yield a unpatterned mixture of different derivatives [14-17].

Typically, stem cells are grown in colonies of variable size and shapes, and the position of a cell within the colony as well as the local cell density has a profound effect on the outcomes of differentiation. Recently, these challenges have been overcome by adopting techniques which control the size and shape of stem cell colonies. In these methods, complimentary patterns of extracellular matrix (ECM) proteins such as laminin and passivating materials that prevent cell and protein adhesion such as poly-L-lysine-grafted-polyethylene glycol (PLL-PEG ) [18] are deposited on the culture surface. Features on the scale of hundreds of microns can be imposed by micro-contact printing where shapes are cast in polydimethylsiloxane (PDMS) elastomer, which are then used as stamps to transfer a pattern onto a slide [19,20]. Alternatives use photolithography, which allows for creating features on the micron scale. The slide is first coated with either the cell attractive or repellant coating, and ultraviolet light shined through a mask is used to burn away the coating in selected regions [18]. When the cells are seeded onto such coverslips, they adhere only where the surface has been coated with ECM proteins and then remain confined to those areas.

The first papers to examine hESC differentiation in micropatterned colonies relied in general on spontaneous rather than morphogen-induced differentiation and did not observe spatial patterning [21,22]. Warmflash et al were the first to consider micropatterns as surrogates for embryonic patterning [23], and it is useful to recapitulate their reasoning and principle results. The mammalian embryo derives from the epiblast, which prior to gastrulation is an apical-basal polarized pseudo-stratified epithelium, sharing a basement membrane with the visceral endoderm. Stem cells grown in microcolonies easily reach densities of $2\text{-}6 \times 10^3$ cells/mm$^2$ similar to the epiblast [24], remain uniform, pluripotent, and display a similar epithelial morphology. The coated surface on which they are grown supplies the basement membrane. Colonies of 0.5-1mm have cell numbers comparable to the mammalian epiblast just prior to gastrulation. Thus, microcolonies are a reasonable platform on which to assay the signals that lead to gastrulation and axis formation in the embryo. When cells are treated with BMP4 ligand, which mimics the gastrulation initiating signal from trophoblast, they initiate a patterning process that allocates cells to all three germ layers along the radial axis of the colony [23]. The outermost cells become trophoblast-like (see discussion of their fate below), the innermost cells differentiate to ectoderm, and rings of mesoderm and endoderm form in between.

While these micropatterned systems represent good models for the epiblastic disc, they do not recapitulate the morphogenesis that occurs in the embryo beginning at gastrulation. They also do not break the radial symmetry of the colony geometry, and the region corresponding to the primitive streak occupies a ring around the colony. Three dimensional culture systems allow for more complex morphogenesis and symmetry breaking, including the formation of apical-basal polarized cysts [25-27], the elongation of the aggregate [28], and the emergence of primitive streak like regions with more natural geometries [26,29]. This more complex morphogenesis comes at a price, as these systems are not quantitatively reproducible in the sense that micropatterned two-dimensional colonies are. All synthetic systems have the potential to break the rigid connections between gastrulation, primitive streaks, and germ layers that we know from embryos. Here we review recent progress on both two and three-dimensional systems that model gastrulation events and their fidelity to the embryo, with a focus on the emergence of both physical structure and cell fate patterns.

**Mathematical Preliminaries**

Subsequent discussion will be enhanced if we impose sharp definitions on terms whose meaning sometimes drifts in biological reviews. Any set of equations describing systems where species can spread by diffusion and undergo a chemical reaction, will fall into the general category of *reaction-diffusion*. It is understood that diffusion in this context is merely a phenomenological approximation to some form of local transfer between cells, the molecular mechanisms are debated and variable. Diffusion does preserve the material being transported, and when this is not the case, say due to molecular traps, one adds an effective decay rate to the system.

We reserve the term *Turing* system to a particular reaction-diffusion system for which the spatially uniform state is unstable and the system evolves towards a periodic pattern, whose wavelength scales as the square root of the diffusion constant divided by a rate [30]. This definition excludes the case where there is a localized source of some activator or preferential signaling at the boundary of a tissue, which then propagates away from the source. The Bicoid gradient in Drosophila is not a Turing system, but does qualify as reaction-diffusion. The pair rule stripes once suggested a Turing mechanism but the reality is the antithesis. A Turing system is capable of spontaneous symmetry breaking, though in development there is almost always some bias that locks the pattern into a particular orientation so that the symmetry is always broken the same way with respect to the body axes. Thus, it is difficult to prove a Turing mechanism for pattern formation purely on the basis of experiments, and arguments in favor typically show the mathematical prerequisites are met and the phenomena resemble what is expected from a Turing model. The so-called activator-inhibitor systems are a particular type of reaction-diffusion model in which a diffusible species activates both its own production and that of a diffusible inhibitor. Under certain conditions, most notably that the inhibitor diffuses faster than the activator, activator-inhibitor systems display Turing properties [31].

A *morphogen* is a signal whose levels can define more than two fates, i.e., we exclude bistable systems from the category of morphogens. Classic examples are Bicoid in Drosophila and Activin/Nodal and BMP in the context of isolated Xenopus animal cap cells [32,33]. Note that the demonstration that a molecule can function as a morphogen in isolated cells does not mean that it necessarily does so in vivo. Putative morphogens including Activin/Nodal, BMP, and Wnt are not static in the vertebrate embryo in contrast to Bicoid [34], and so the interpretation of these signals can be complex. Nonetheless, they can still convey positional information i.e., distance from a defined source. For example, if the signal transduction pathway is adaptive, that is returns to its pre-stimulus baseline after a step increase in morphogen concentration, then its quantitative response is proportional to the time rate of change of the morphogen. If a morphogen turns on at a defined time and spreads, points near the source will experience a more abrupt change than points further away. Thus, a dynamic signal can convey positional information to an adaptive receiver. This is all easy to demonstrate mathematically, and the sensitivity of signaling outputs to the rate of change of TGFb ligands has been shown in a cell culture system [35].

Some signals, notably WNT [36], operate at short distances or only by cell contacts and alone cannot coordinately pattern an embryo with a diameter of hundreds of microns. Nonetheless, the inhibitors are often longer range (as required for an activator-inhibitor Turing system) and can impose a pattern on a background of constant activator production [31]. Thus, it's of interest to study the movement of the inhibitors and micropatterned colonies could serve as an attractive platform for evaluating their range and mechanisms of action. One potential objection is that in cell culture secreted signals may escape into the bulk media and therefore not be relevant to patterning. Experimentally, this appears not to be the case, as knockdown or knockout of secreted inhibitors has clear patterning phenotypes in micropatterned colonies [23,24]. In cell culture, it is generally true that some secreted signals escape into the media and are homogeneous, nevertheless, autocrine signaling can occur even when the conditioned media transferred to naïve cells does not elicit paracrine signaling arguing that local signaling is still possible in cell culture.

To understand the distribution of inhibitors on micropatterns, a related effect should be noted. Assume an inhibitor is made uniformly, secreted, and adsorbed back onto the cell layer (perhaps to be endocytosed, but for whatever reason remains attached). If the inhibitor is released a distance $z_0$ above the disk away from the edges, then in a time of order of $z_0^2/D$ all the inhibitor will be readsorbed on the surface, where D is the diffusion constant in the media. At the edge of the colony the inhibitor can mix into the volume. The net result is that the profile of inhibitor can be described by two-dimensional diffusion within the layer (either directly cell to cell, or via secretion and local uptake from the media), and fixed at a low value at the edge [24]. This will restrict activity of the activator to the colony edge with a range depending on the concentration of supplied activator.

**Colony Architecture:**

In both the embryo and synthetic systems, the morphology of cells and tissues has a large impact on how signals are transmitted and ultimately how fates are acquired. We thus consider common physical aspects of stem cell systems before turning to fate determination.

*Growth in Two Dimensional Micropatterns*

The apical-basal structure of micropatterned colonies in the pluripotent state was investigated in ref [24]. They show that the apical tight junction marker ZO-1 and the centrioles were positioned on the apical side of the nucleus. In common with other polarized epithelia [37], both the Activin/Nodal and BMP receptors are localized to the baso-lateral sides of the cells. This is more pronounced at high cell densities with the result that colonies become insensitive to apically applied morphogens. At the colony boundaries, however, the apical-basal axis becomes more radial perhaps associated with the stress fibers one finds there [38], with the result that the receptors remain apically exposed. The most compelling data for the receptor polarization arise from confluent cell colonies grown on filters. These are ~10μm thick transparent membranes with 10-200 ~0.4 μm pores per cell. They are sealed into wells so that different media can be placed on the two sides. The strong asymmetry in response between apically and basally applied BMP or Activin ligands argues for basolateral receptor localization. There are anecdotal observations that colony edges tended to differentiate before the bulk, but that was not connected with receptor occlusion. Growth on filters with TGFβ ligands supplied from below is a simple technique to insure uniform application of cytokines when uniform signaling is desired. It is yet to be widely adapted in the stem cell field.

Any of the technologies used to make micropatterns, can also make arbitrary shapes. This fact was exploited by Blin et al. to make lozenge-shaped domains and examine the effect of the corners on cell fate [39]. Growing mouse ESCs under pluripotent conditions, they observed some spontaneous differentiation to Bra+ cells, and controlled the fraction of such cells by adjusting colony density prior to replating. They observed the Bra+ cells preferentially moved to the corners. This is consistent with old ideas that tissues behave as if endowed with a surface tension, so in this case we would infer that the Bra+ cells optimize their contact with media in preference to the undifferentiated cells by occupying the corners. Since there are no supplied morphogens or patterns of signaling, and the colonies appear to be somewhat layered from the start, the embryological relevance is unclear.

*Three-Dimensional Culture Systems*

There is a long history of papers tracing the influence of extracellular-matrix on cancer as regards its chemical composition, mechanical properties, and dimensionality. The reconstitution of breast acinar networks from normal and cancerous endothelial cells is particularly revealing about the importance of the 3D physical environment of the cells [40]. These methods have slowly found their way to the stem cell field.

An interesting illustration of their potential is described in ref [27]. Cells are first seeded on a soft matrigel layer, allowed to form colonies for a day, and then embedded in a dilute matrigel solution, that favors the formation of closed epithelial cysts. Presumably the matrigel solution encourages cells to place their basal sides out, but nothing is known about the transition intermediate between the layer and the cyst, perhaps it resembles a neural rosette with the apical surfaces grouped into a circle and the basal sides radially extended. A combination of matrigel in the media and soft substrate for growth are both required for the colony to spontaneously differentiate to squamous epithelial morphology and display a gene signature indicative of human amnion. Three-dimensional cysts form with either a soft culture substrate and standard growth media or on a hard surface with matrigel added to the culture media, but they remain columnar and pluripotent. Thus, cell contacts and the physical environment of the colony can have a profound effect on cell fates in the absence of supplied morphogens. We still know very little about outcomes when morphogens and the physical environment compete, or the extent to which cells in a cyst will reconstruct their own basal membrane de-novo once some global cue establishes their collective polarity.

For the purposes of massively expanding human stem cell numbers Lei et al used a PEG based hydrogel that solidified when the temperature was raised to 37C, that together with a chemically defined growth media, allowed single cells to expand to ~350μ diameter balls that remained fully pluripotent [41]. This should be contrasted with large 3D aggregates, called embryoid bodies, made from cells first grown on surfaces and then placed in suspension in differentiation media which differentiate in a mostly disorganized fashion and sometimes show apoptosis in the center [42].

For 3D stem cell culture with outcomes more relevant to the embryo, we have mostly data from mouse. In conjunction with their study of how the inner cell mass reorganizes to form the epiblast and amniotic cavity, the Zernicka-Goetz lab put mouse ESCs directly into matrigel [25]. They formed a polarized epithelial cyst once more than a few cells were present, whose formation required the ECM components of matrigel. It is not yet clear how long the cysts remain pluripotent under these conditions, and whether by measures of gene expression the cells successively transit from their inner cell mass state to the pluripotent epiblast state.

Much larger cysts, again starting from mouse ESC, were induced by a neural differentiation protocol to form a dense polarized epithelium resembling the neural plate [43]. They respond to signals that regulated their fates along the anterior-posterior axis. In common with a neural epithelium, cells move to the apical surface prior to division. A similar culture has not yet been reported for human cells, but if the patterning mechanism follows that in-vivo, it could prove to be a useful assay for the interaction of SHH and BMP signaling.

The extra cellular matrix is generally consigned to a supporting role in morphogenesis, necessary but otherwise ignored. In a follow up to [43], Ranga et al. used synthetic hydrogels where they could control the ECM components (as well as mechanics) and systematically screened for the properties of the neural-cysts and their propensity for spontaneous DV axis formation [44]. Their system revealed the generation of ECM by the cyst itself and how the basement membrane remodeled as the cyst grows. This may prove to be a feasible route to resolving how the specific components of the ECM contribute to morphogenesis.

Mouse is an appealing system in which to explore co-culture of different cell types since stable cell lines exist for the lineages that derive from the blastocyst: primitive endoderm, trophoblast, and the standard ICM-derived ESC. In Harrison et al, trophoblast cell colonies were mixed with the cultures that generated the epiblast cysts, and led to the formation of structures resembling the egg cylinder with the trophoblast ball capping the epiblast epithelial shell. The two cell populations established a common luminal compartment as in the embryo, and then showed asymmetric expression of Bra and Wnt activity [26]. Cell fate patterning in this system is discussed below.

There is a considerable literature on the influence of substrate stiffness on the fate of stem cell colonies undergoing spontaneous differentiation[45]. A recent paper using hESC shows that a soft substrate can enhance a mesoderm induction [46]. In this case, cells on soft substrates preferred E-Cadherin dependent cell-cell contacts to integrin-dependent contacts with the culture surface. This led to upregulation of β-Catenin and greater sensitivity to a mesoderm induction protocol. However, this protocol did not include Wnt, and they showed that cells cultured on stiff surfaces and supplemented with Wnt, gave results similar to soft surfaces. Since multiple papers produce various mesoderm derived fates on glass with good efficiency [12,47], we conclude that substantial doses of morphogens can override the effects of mechanics. An earlier paper however demonstrated enhanced yields of neural progenitors when subject to dual smad inhibition on soft substrates as compared to stiff ones [48]. More generally, even if supplied morphogens can override the effects of mechanics in culture, mechanics may still play an important role in influencing differentiation outcomes at physiological concentrations in vivo.

**Spatial patterning of cell fates**

As embryogenesis proceeds, the cells of the embryo differentiate to appropriate fates depending on their spatial position. A complex network of ligands and their inhibitors is used to instruct these fate decisions. In vivo, these decisions are entwined with the processes of growth, cell division, and morphogenesis making quantitative study difficult. Moreover, while it is relatively straightforward to determine the patterns of gene expression for the mRNAs encoding the ligands and inhibitors, determining the spatial distributions of the proteins themselves as well as the signaling responses has proved much more difficult. Studying these processes in stem cells offers a potential alternative as imaging is considerably more straightforward in stem cell cultures than in mammalian embryos, and ligands can be applied in a controlled fashion making it possible to determine quantitative dynamic input-output relationships for each signaling pathway. In this section, we focus on recent progress studying cell fate patterning associated with gastrulation in 2D and 3D cultures of mouse and human ESCs.

*Micropatterned two dimensional culture systems.*
The signals governing patterning within micropatterned colonies treated with BMP4 are the same as those governing gastrulation in the mouse embryo [Arnold-Robertson 2009]. The externally supplied BMP4 activates transcription of Wnt ligands which in turn activate Nodal. Both Wnt and Nodal signals are required for the differentiation of the mesendoderm. The ligands themselves are not sufficient to generate the spatial pattern, and the Nodal inhibitors Lefty and

Cereberus restrict the mesoderm to the rings. Without these inhibitors, the mesoderm will spread to fill the colony. In the absence of either Nodal or Wnt signals, mesendodermal fates are lost and the colony is divided between trophectodermal fates at the colony border and ectodermal fates at the center [23].

While the patterning of the germ layers is similar to the embryo, the differentiation of the outer cells from epiblast-like hESCs to trophoectoderm is quite different from the situation in vivo where the epiblast derives from the ICM only after it has split from the trophectodermal lineages. As a consequence, the identity of these cells has remained controversial, with some suggesting that they represent extraembryonic mesoderm rather than trophoblast [15]. More recently, a substantial amount of data has been obtained showing similar transcriptional profiles, hormone secretion, and physiological responses between BMP4 differentiated hESCs and trophoblast [49-52]. Nonetheless, data showing that these cells can actually function in vivo are lacking. Thus, it remains unclear whether these cells represent true trophoblast that are differentiated by a different path than their in vivo counterparts, a different but molecularly similar cell type, or possibly a culture artifact which bears a resemblance to trophoblast, but does not correspond to any cell occurring in the embryo.

Whatever the precise identity of these cells, it is clear that their differentiation is dependent on BMP signaling. In micropatterns, an initially broad response to the added BMP ligand is refined over time so that only the trophectoderm-like cells at the border show sustained BMP signaling[23]. In experiments with sparsely seeded cells in standard culture, it was shown that this sustained BMP response is required for cells to adopt this fate [53], and terminating signaling early also prevented differentiation. During patterning in larger colonies, the restriction of these signals to the colony border is dependent on two factors, the prepattern in apical-basal receptor localization discussed above [24], and inhibition by the secreted inhibitor Noggin. As discussed above, uniform production of Noggin within the colony together with diffusion are sufficient to create a gradient with the highest levels of Noggin at the colony center and the lowest levels at the edge. As Noggin is a direct BMP target in hESCs, as well as a Nodal target, it is likely that the patterns of BMP and Nodal signaling induce patterns of Noggin expression and that these play a role in shaping the resulting cell fate patterns.

As noted above, a cascade of signaling events is responsible for initiating the gastrulation-like processes in micropatterned colonies. The exogenously supplied BMP activates Wnt signaling which in turn activates Nodal. Both extracellular Nodal and Wnt inhibitors are required for limiting the spread of these signaling activities and the resulting mesendoderm differentiation. The architecture of these signaling circuits is reminiscent of the theoretically well-studied activator-inhibitor systems originally proposed by Meinhardt (reviewed in [31]) which are examples of Turing systems (see discussion above). Nodal both activates itself and its extracellular inhibitors Lefty1/2 and Cerberus. Similarly, Wnt signaling activates both the Wnt3 ligand and its extracellular inhibitor Dkk1. Nodal and its inhibitors lefty have also been proposed to act as these type of Turing systems in other contexts [54,55].

If Wnt and Nodal do indeed function as Turing systems in this context, they would be capable of generating similar patterns even in the absence of induction by BMP at the edge but these would be variable within the colony. That is, a stripe or patch of high-signaling cells would form stochastically at a particular position and inhibit further signaling and mesendoderm differentiation in the region around it. The function of the upstream BMP signaling is to bias this process so that the pattern is always the same from the edge of the colony inward. Similar mechanisms have been suggested to take place in other patterning systems. For example, the ventral neural tube is patterned under the control of the morphogen Sonic hedgehog (Shh). Shh expressed from the neural tube itself could create patterns in a self-organized fashion, but which side of the neural tube adopted a ventral fate would be random. Shh from the notochord, which lies ventral to the neural tube can bias the patterning process so that the ventral side of the neural patterning always aligns with the ventral side of the embryo [3].

Taken together, these results suggest a two-step patterning process. First, a combination of high Noggin concentrations and inaccessible receptors at the center of the colony restricts the response to exogenous BMP4 to the colony edge. BMP4 then activates two potential Turing systems, Wnt signaling and Nodal signaling which position stripes of these activities to the primitive streak like region where mesendoderm differentiation occurs. The finding that BMP signaling activates the inhibitor Noggin [24] raises the possibility that BMP-Noggin also acts as a Turing system in patterning. In the future, it will be interesting to rigorously examine this possibility as well as the Wnt and Nodal patterning systems to better understand the relationships between these three and the patterns they generate.

Recently, another study has confirmed these experimental findings but proposed an alternative explanation for the observed pattern of cell fates in micropatterned colonies [56]. Tewary et al. grew micropatterned colonies in a defined medium containing recombinant Nodal, differentiated these by adding BMP4, and found identical patterns to those in the micropatterning studies reviewed above. They proposed that a Turing system involving BMP4 and Noggin creates a gradient of BMP signaling as reflected in the activated signal transducer Smad1. The levels of pSmad1 are then proposed to determine cell fates in a concentration-dependent manner. As evidence for this model, they show that in smaller colonies, which typically differentiate entirely to the trophoectodermal fates found at the edges of large colonies, expression of mesoderm and ectodermal are induced by lower doses of BMP.

A number of experimental observations argue against this model. Experiments on BMP signaling in hESCs both in the context of micropatterned culture and in standard culture suggest that BMP cannot function as a classic morphogen in this context. First, while pSmad1 is highest at the colony edge, there isn't a clear gradient of activity. In fact, the distribution of pSmad1 can be effectively modeled as a binary distribution with cells either on or off [24]. Three different studies show that pSmad1 is restricted to within about 100μm of the edge of the colony [23,24,56], which is too narrow a range to pattern all the cell fates within the colony in a concentration-dependent manner. The gradient is broader earlier in patterning, possibly suggesting a duration-dependent interpretation of BMP signaling with longer exposure needed for trophoectodermal than mesodermal fates, however, this is contradicted by the absolute requirement for both Nodal and Wnt signaling in forming the mesendoderm in these colonies [23,56]. Finally, experiments examining the dose-dependent response to BMP4 in very small colonies, which lack secondary signals, show that only a single fate is generated. That is, cells switch from pluripotent to trophectodermal fates above a threshold concentration without any alternative fates generated, supporting a binary model of cell fate decisions induced by BMP. In larger colonies, mesodermal fates are generated but these require secondary signals, and consistently, are only observed at particular cell densities [57].

Further, the evidence in Tewary et al are not consistent with a BMP4 and Noggin forming a Turing system in the sense defined above. As noted above, Turing systems generate self-organized patterns with fixed length scales determined by diffusion and decay constants. If BMP-Noggin formed such a system, the role of the exogenous BMP4 would be to trigger the formation of these self-organized patterns with a bias towards the edge, and the resulting patterns would be independent of BMP4 dose once the self-organizing system had been activated. The fact that the patterns can be rescued with lower BMP4 doses suggests that such a Turing system is not operating. If instead, there were a gradient of Noggin that is highest in the center, then the range over which BMP4 could overcome the Noggin repression would be dependent on the BMP4 dose. Further, the existence of doses of BMP4 that do not show the BMP4 dependent CDX2 fates at the colony border but do show mesendoderm differentiation is also consistent with the two step model proposed above. Experiments show that CDX2 fates require sustained high levels BMP signaling, significantly beyond the times shown to be required to activate Nodal and Wnt signals [57]. Thus, at some doses, the level or duration of BMP signaling will not be high enough to give rise to CDX2 fate but will be sufficient to activate the Wnt and Nodal patterning systems giving rise to the mesendodermal fates at the colony edge.

During cell fate patterning, coherent territories of a single fate are generated, and signals between the cells of the territory may be required for differentiation. John Gurdon originally demonstrated that groups of Xenopus animal cap cells, but not individual cells, are induced to form muscle by interaction with vegetal cells [58]. Positive feedback in which signaling pathway activity enhances transcription of the genes encoding the pathway ligands has been proposed to generate coherent signaling and fate responses within a group of cells [59], while negative feedback might be required to limit the extent of this territory, and create cells fate patterns as discussed above [60].

Recently, micropatterning approaches have been used to investigate the mechanisms underlying these phenomena at the single cell level [57]. When hESCs grown in small colonies of 1-8 cells are treated with BMP4, cells within each colony coordinate their response so that at intermediate doses where both pluripotent and trophoectodermal fates are present, each colony is typically composed of entirely CDX2+ or SOX2+ positive fates. These trends are strengthened as the colony size increases. Further, this trend towards uniformity within the colony reinforces the fates instructed by exogenously supplied signals, so that compared to smaller colonies, those with four or more cells retain pluripotency better in pluripotency supporting media and differentiate more sensitively and homogenously in response to BMP4. At the level of signaling, live cell imaging showed that larger colonies are better able to sustain the response to the BMP signal and therefore differentiate more homogenously. Smaller colonies show more variable signaling and differentiation. Correlating fates with signaling at the level of single cells shows that it is the cells with sustained signaling that differentiate to the trophectodermal fates. Positive feedback between BMP signaling and transcription of BMP ligands is a plausible molecular mechanism for these observations, but this remains to be tested.

*Three dimensional culture systems*
Early embryonic events have also been investigated in three-dimensional cultures of mESCs. In initial experiments, it was shown that embryoid bodies made from mESCs show spontaneous polarization of a Wnt signaling reporter and mesodermal gene expression suggestive of an anterior-posterior axis [61]. It was also shown that the hierarchy of signaling from BMP to Wnt to Nodal is preserved, so that treatment with any of BMP, Wnt, or Activin can lead to activation of polarized Wnt signaling in these aggregates, but that BMP-inhibition only blocks the polarization induced by BMP. Wnt and Activin are downstream of BMP and so activate polarized Wnt activity in a BMP-independent fashion. More recently, it was shown that when these aggregates are made from relatively small numbers of cells, the polarization is accompanied by elongation along this axis, with the posterior markers on one end. This effect can be enhanced by Wnt activation during a particular period in the culture [28]. Moreover, it was observed in some aggregates that neural markers such as Sox1 and Sox2 are not expressed opposite the region of Bra expression on the long axis of the aggregate but instead on the shorter axis (Figure 1A), and it was suggested that this represents a second axis in the aggregate, akin to the DV axis of the embryo [62].

While intriguing, further experiments will be required to support these claims. First, outside the context of the embryo, localized expression of germ layer markers such as Bra or Sox2 may result from the process of germ layer differentiation either under the spatial control of ligands or through more stochastic processes followed by cell sorting (as in [39]), rather than the formation of an axis equivalent to the AP axis of the embryo. Additional markers specific to particular AP positions such as Otx2 for anterior fates or particular Hox genes for more posterior ones could support these conclusions. It is also possible that more elaborate protocols will be required to define the AP position. For example, it was recently shown that neural/mesodermal progenitors can be maintained in a combination of Wnt and FGF signaling and during this time acquire a progressively more posterior identity as defined by Hox gene expression. Treatment with retinoic acid at any time during this protocol induces differentiation to neural fates and freezes the AP identity of the cells [10]. The claim of two independent axes requires multiple markers to be assayed and simultaneously visualized. In the absence of this, it is equally possible that axial elongation and AP axis formation can be decoupled in aggregates so that the Bra-Sox1/2 axis apparent in Figure 1A corresponds to an AP axis or, as noted above, to germ layer differentiation without a clear correspondence to one of the major body axes. The DV axis would most clearly be demonstrated by visualizing ventral and dorsal fates within the same germ layer, for example, neural and epidermal fates within the ectoderm. These issues are also complicated by the variability seen within aggregates. All aggregates form a long axis with Bra and Wnt signaling on one end, allowing for quantification of these markers relative to this axis, but other aspects such as the Sox2 expression appear variable making it difficult to have an external reference by which all markers can be compared.

As discussed above, Harrison et al [26] developed a three-dimensional culture system which combines mESCs and trophoblast stem cells (TSCs) into a structure called an ETS embryo (for ESC and TSC derived embryo). In addition to recapitulating egg cylinder stage morphogenesis, ETS embryos also show asymmetric expression of primitive streak markers such as Brachyury and germ cell markers such as Stella, Figure 1C. Thus, these embryos have two orthogonal axes, the proximal-distal axis, defined by the relative position of the ESCs and TSCs, and an AP like axis in which the positioning of Brachyury and germ cell markers defines the posterior side. The development of the AP axis is particularly interesting as it occurs in the absence of the visceral endoderm, while in vivo, secreted signals from the anterior visceral endoderm are required to position the primitive streak in the posterior of the embryo [6]. An attractive model is that the generation of the primitive streak is under the control of a Turing system so that it stochastically forms on one side of the ETS embryo, and then the longer-range inhibitors prevent further Wnt/Nodal signaling and primitive streak formation on the opposite side. If this model is correct, an open question is what prevents similar mechanisms from operating in real embryos lacking the secreted inhibitors in the AVE [6], in embryos in which the AVE does not form [63,64], or in the micropatterned human ESC colonies discussed above [23]. In the former case, multiple primitive streaks form, while in the later two cases, the radial symmetry of the embryo or colony is never broken resulting a ring of mesodermal differentiation rather than a streak on one side. The patterning by sorting rather than morphogens is not excluded in this system either.

Both these 3D systems lose essential aspects of in-vivo gastrulation. A cell aggregate does not undergo the epithelial to mesenchyml transition (EMT), which is a necessary step in primitive streak formation. The ETS embryos do not have a well characterized EMT or a mesenchyml layer covering the remaining epiblast epithelium. Many protocols exist to make mesendo derivatives from stem cells without obvious intermediate spatial organization. Would mixtures of cells fated to different germ layers sort and look so different from synthetic systems?

An example of a self-patterning three-dimensional system in human is the amniotic cysts discussed above. In most cases, these create relatively homogeneous aggregates of amniotic ectoderm, however, it was recently shown that in a minority of cases, polarized cysts form consisting of an amniotic half and an epiblastic half in which the epiblastic half retains its columnar epithealial morphology while the amniotic half differentiates to a squamous epithelium expressing markers of amnion such GATA3 and CDX2 [29](Figure 1B). As with the fully differentiated cysts, the differentiation of the amniotic half in polarized cysts requires BMP signaling which automously becomes asymmetric in the cyst. It is hypothesized that in the polarized cysts, BMP induction of BMP inhibitors limits the spread of amniotic differentiation and allows for the stable retention of epiblast fates in half the aggregate. This hypothesis remains to be proven, and, in any event, it remains unclear what distinguishes the fully differentiated cysts where the BMP-mediated differentiation spreads to the entire aggregate and the polarized cysts in which it is limited. Interestingly, in polarized cysts, a primitive streak like region often develops from the epiblastic part, however, it remains unknown whether the amniotic half of the cyst plays a role in inducing this event, as the extraembryonic tissue does in vivo, or whether it arises spontaneously from the epiblast cells. If the amnion plays a role, the primitive-streak like region should initiate at the border between the epiblast and amnion cells and extend from there towards the center of the epiblast region, and it will be interesting to determine whether this is the case.

Finally, many examples of systems that undergo patterning and morphogenesis and model the development of particular organs have recently been developed (reviewed in [2,3,65]). One of the most relevant to early development are the neural cysts discussed above [43]. In addition to the morphogenesis discussed above, they also represent an interesting in vitro system for studying cell fates within the developing neural tube. Cyst grown in neural induction conditions were uniformly anterior and dorsal, and could be ventralized through activation of the Shh hedgehog pathway. Interestingly, treatment with RA induced more posterior fates and also led to spontaneous dorsal-ventral patterning as assayed by sonic hedgehog (SHH) in the putative floor plate and several early motor neuron fate markers in their correct relative positions.

**Conclusions**

The stem cell systems reviewed here represent promising avenues for making progress on difficult problems in mammalian embryogenesis. To date, most work has shown that these systems recapitulate already known features of mammalian embryogenesis such as the cascade of signaling from BMP to Wnt to Nodal, however, new insights which may be applicable to the embryo are also beginning to emerge. One example is the role of both secreted inhibitors and receptor localization in restricting the response of the epiblast to BMP. In vivo, it is possible that localizing the receptors to the basal side of the embryo both restricts signaling to the epiblast – extraembryonic (trophoblast or amnion) boundary where this localization breaks down. It further prevents signaling from cavity, which is apical to the cells, from globally initiating gastrulation. The use of filter systems and micropatterned colonies has begun to unravel these interactions in culture, and it will be important to test their relevance in vivo in the future.

The culture systems can also be combined with mathematical modeling to investigate fundamental issues of symmetry breaking in development. In this regard, while several experiments suggest that Wnt-Dkk or Nodal-Lefty function as activator-inhibitor systems to generate Turing patterns, it is notable that they do not break the azimuthal symmetry of the colonies but instead generate rings of primitive streak formation. This is in contrast to the situation in vivo in which the primitive streak only occupies the posterior side of the embryo. It is possible that by treating these colonies with high levels of BMP, which they strongly respond to on the entire perimeter, they are constrained to adopt azimuthally symmetric organizations of signaling and fate. More natural ways of inducing the gastrulation might reveal whether the cells are intrinsically capable of breaking this symmetry or whether interactions with extraembryonic tissues which are lacking in these culture systems are required.

**Figure Caption**
Examples of self-patterning in 3D. (a) An embroid body from mESC stained for Bra (red) and Sox2 (blue), and with Sox1::GFP in green [62] (b) A structurally asymmetric amniotic cyst from hESC, with a thick pseudostratified epithelium on one side and a thin amnion layer on the other, stained for nuclei (blue), βCAT (green) and E-CAD (red)[29]. (c) The juxtaposition of mouse trophoblast cells (top) with an epiblast epithelium (bottom) results in an incipient primitive streak (right) breaking the azimuthal symmetry in the ring where the two types of cells are in contact, nuclei (red), Oct4 (blue), Stella-GFP (green) [26]

**Acknowledgement**
We thank Idse Heemskerk, Alfonso Martinez-Arias, and Mijo Simunovic for comments on an earlier draft of this review. EDS was supported in part by NSF grant PHY 1502151, and AW by XXX

**Reference**

[1]   Heemskerk, I.; Warmflash, A. Pluripotent Stem Cells as a Model for Embryonic Patterning: From Signaling Dynamics to Spatial Organization in a Dish. *Dev Dyn*, **2016**, *245*, 976–990.
[2]   Simunovic, M.; Brivanlou, A.H. Embryoids, Organoids and Gastruloids: New Approaches to Understanding Embryogenesis. *Development*, **2017**, *144*, 976–985.
[3]   Turner, D.A.; Baillie-Johnson, P.; Martinez Arias, A. Organoids and the Genetically Encoded Self-Assembly of Embryonic Stem Cells. *Bioessays*, **2016**, *38*, 181–191.
[4]   Sasai, Y. Cytosystems Dynamics in Self-Organization of Tissue Architecture. *Nature*, **2013**, *493*, 318–326.
[5]   Arnold, S.J.; Robertson, E.J. Making a Commitment: Cell Lineage Allocation and Axis Patterning in the Early Mouse Embryo. *Nat Rev Mol Cell Biol*, **2009**, *10*, 91–103.
[6]   Perea-Gomez, A.; Vella, F.D.J.; Shawlot, W.; Oulad-Abdelghani, M.; Chazaud, C.; Meno, C.; Pfister, V.; Chen, L.; Robertson, E.; Hamada, H.; Behringer, R.R.; Ang, S.-L. Nodal Antagonists in the Anterior Visceral Endoderm Prevent the Formation of Multiple

Primitive Streaks. *Dev Cell*, **2002**, *3*, 745–756.
[7] Ben-Haim, N.; Lu, C.; Guzman-Ayala, M.; Pescatore, L.; Mesnard, D.; Bischofberger, M.; Naef, F.; Robertson, E.J.; Constam, D.B. The Nodal Precursor Acting via Activin Receptors Induces Mesoderm by Maintaining a Source of Its Convertases and BMP4. *Dev Cell*, **2006**, *11*, 313–323.
[8] Dunn, N.R.; Vincent, S.D.; Oxburgh, L.; Robertson, E.J.; Bikoff, E.K. Combinatorial Activities of Smad2 and Smad3 Regulate Mesoderm Formation and Patterning in the Mouse Embryo. *Development*, **2004**, *131*, 1717–1728.
[9] Chambers, S.M.; Fasano, C.A.; Papapetrou, E.P.; Tomishima, M.; Sadelain, M.; Studer, L. Highly Efficient Neural Conversion of Human ES and iPS Cells by Dual Inhibition of SMAD Signaling. *Nat Biotechnol*, **2009**, *27*, 275–280.
[10] Lippmann, E.S.; Williams, C.E.; Ruhl, D.A.; Estevez-Silva, M.C.; Chapman, E.R.; Coon, J.J.; Ashton, R.S. Deterministic HOX Patterning in Human Pluripotent Stem Cell-Derived Neuroectoderm. *Stem Cell Reports*, **2015**, *4*, 632–644.
[11] Teo, A.K.K.; Arnold, S.J.; Trotter, M.W.B.; Brown, S.; Ang, L.T.; Chng, Z.; Robertson, E.J.; Dunn, N.R.; Vallier, L. Pluripotency Factors Regulate Definitive Endoderm Specification Through Eomesodermin. *Genes Dev*, **2011**, *25*, 238–250.
[12] Loh, K.M.; Chen, A.; Koh, P.W.; Deng, T.Z.; Sinha, R.; Tsai, J.M.; Barkal, A.A.; Shen, K.Y.; Jain, R.; Morganti, R.M.; Shyh-Chang, N.; Fernhoff, N.B.; George, B.M.; Wernig, G.; Salomon, R.E.A.; Chen, Z.; Vogel, H.; Epstein, J.A.; Kundaje, A.; Talbot, W.S.; Beachy, P.A.; Ang, L.T.; Weissman, I.L. Mapping the Pairwise Choices Leading From Pluripotency to Human Bone, Heart, and Other Mesoderm Cell Types. *Cell*, **2016**, *166*, 451–467.
[13] Pagliuca, F.W.; Millman, J.R.; Gürtler, M.; Segel, M.; Van Dervort, A.; Ryu, J.H.; Peterson, Q.P.; Greiner, D.; Melton, D.A. Generation of Functional Human Pancreatic B Cells in Vitro. *Cell*, **2014**, *159*, 428–439.
[14] Tang, C.; Ardehali, R.; Rinkevich, Y.; Seita, J.; Lee, A.S.; Mosley, A.R.; Drukker, M.; Weissman, I.L.; Soen, Y. Isolation of Primitive Endoderm, Mesoderm, Vascular Endothelial and Trophoblast Progenitors From Human Pluripotent Stem Cells. *Nat Biotechnol*, **2012**, *30*, 531–542.
[15] Bernardo, A.S.; Faial, T.; Gardner, L.; Niakan, K.K.; Ortmann, D.; Senner, C.E.; Callery, E.M.; Trotter, M.W.; Hemberger, M.; Smith, J.C.; Bardwell, L.; Moffett, A.; Pedersen, R.A. BRACHYURY and CDX2 Mediate BMP-Induced Differentiation of Human and Mouse Pluripotent Stem Cells Into Embryonic and Extraembryonic Lineages. *Cell Stem Cell*, **2011**, *9*, 144–155.
[16] Yu, P.; Pan, G.; Yu, J.; Thomson, J.A. FGF2 Sustains NANOG and Switches the Outcome of BMP4-Induced Human Embryonic Stem Cell Differentiation. *Cell Stem Cell*, **2011**, *8*, 326–334.
[17] Xu, R.-H.; Chen, X.; Li, D.S.; Li, R.; Addicks, G.C.; Glennon, C.; Zwaka, T.P.; Thomson, J.A. BMP4 Initiates Human Embryonic Stem Cell Differentiation to Trophoblast. *Nat Biotechnol*, **2002**, *20*, 1261–1264.
[18] Azioune, A.; Storch, M.; Bornens, M.; Théry, M.; Piel, M. Simple and Rapid Process for Single Cell Micro-Patterning. **2009**, *9*, 1640–1642.
[19] Qin, D.; Xia, Y.; Whitesides, G.M. Soft Lithography for Micro- and Nanoscale Patterning. *Nat Protoc*, **2010**, *5*, 491–502.
[20] Théry, M.; Piel, M. Adhesive Micropatterns for Cells: a Microcontact Printing Protocol.

*Cold Spring Harb Protoc*, **2009**, *2009*, pdb.prot5255–pdb.prot5255.

[21] Peerani, R.; Rao, B.M.; Bauwens, C.; Yin, T.; Wood, G.A.; Nagy, A.; Kumacheva, E.; Zandstra, P.W. Niche-Mediated Control of Human Embryonic Stem Cell Self-Renewal and Differentiation. *EMBO J*, **2007**, *26*, 4744–4755.

[22] Bauwens, C.L.; Peerani, R.; Niebruegge, S.; Woodhouse, K.A.; Kumacheva, E.; Husain, M.; Zandstra, P.W. Control of Human Embryonic Stem Cell Colony and Aggregate Size Heterogeneity Influences Differentiation Trajectories. *Stem Cells*, **2008**, *26*, 2300–2310.

[23] Warmflash, A.; Sorre, B.; Etoc, F.; Siggia, E.D.; Brivanlou, A.H. A Method to Recapitulate Early Embryonic Spatial Patterning in Human Embryonic Stem Cells. *Nat Meth*, **2014**, *11*, 847–854.

[24] Etoc, F.; Metzger, J.; Ruzo, A.; Kirst, C.; Yoney, A.; Ozair, M.Z.; Brivanlou, A.H.; Siggia, E.D. A Balance Between Secreted Inhibitors and Edge Sensing Controls Gastruloid Self-Organization. *Dev Cell*, **2016**, *39*, 302–315.

[25] Bedzhov, I.; Zernicka-Goetz, M. Self-Organizing Properties of Mouse Pluripotent Cells Initiate Morphogenesis Upon Implantation. *Cell*, **2014**, *156*, 1032–1044.

[26] Harrison, S.E.; Sozen, B.; Christodoulou, N.; Kyprianou, C.; Zernicka-Goetz, M. Assembly of Embryonic and Extra-Embryonic Stem Cells to Mimic Embryogenesis in Vitro. *Science*, **2017**, eaal1810.

[27] Shao, Y.; Taniguchi, K.; Gurdziel, K.; Townshend, R.F.; Xue, X.; Yong, K.M.A.; Sang, J.; Spence, J.R.; Gumucio, D.L.; Fu, J. Self-Organized Amniogenesis by Human Pluripotent Stem Cells in a Biomimetic Implantation-Like Niche. *Nat Mater*, **2016**, 1–9.

[28] van den Brink, S.C.; Baillie-Johnson, P.; Balayo, T.; Hadjantonakis, A.-K.; Nowotschin, S.; Turner, D.A.; Martinez Arias, A. Symmetry Breaking, Germ Layer Specification and Axial Organisation in Aggregates of Mouse Embryonic Stem Cells. *Development*, **2014**, *141*, 4231–4242.

[29] Shao, Y.; Taniguchi, K.; Townshend, R.F.; Miki, T.; Gumucio, D.L.; Fu, J. A Pluripotent Stem Cell-Based Model for Post-Implantation Human Amniotic Sac Development. *Nat Commun*, **2017**, 1–14.

[30] Turing, A. The Chemical Basis of Morphogenesis. *Philosophical Transactions of the Royal Society of London. Series BBiological Sciences*, **1952**, *237*, 37–72.

[31] Meinhardt, H. Models of Biological Pattern Formation: From Elementary Steps to the Organization of Embryonic Axes. *Curr. Top. Dev. Biol.*, **2008**, *81*, 1–63.

[32] Green, J.B.; New, H.V.; Smith, J.C. Responses of Embryonic Xenopus Cells to Activin and FGF Are Separated by Multiple Dose Thresholds and Correspond to Distinct Axes of the Mesoderm. *Cell*, **1992**, *71*, 731–739.

[33] Wilson, P.A.; Lagna, G.; Suzuki, A.; Hemmati-Brivanlou, A. Concentration-Dependent Patterning of the Xenopus Ectoderm by BMP4 and Its Signal Transducer Smad1. *Development*, **1997**, *124*, 3177–3184.

[34] Schohl, A.; Fagotto, F. Beta-Catenin, MAPK and Smad Signaling During Early Xenopus Development. *Development*, **2002**, *129*, 37–52.

[35] Sorre, B.; Warmflash, A.; Brivanlou, A.H.; Siggia, E.D. Encoding of Temporal Signals by the TGF-B Pathway and Implications for Embryonic Patterning. *Dev Cell*, **2014**, *30*, 334–342.

[36] Farin, H.F.; Jordens, I.; Mosa, M.H.; Basak, O.; Korving, J.; Tauriello, D.V.F.; de Punder, K.; Angers, S.; Peters, P.J.; Maurice, M.M.; Clevers, H. Visualization of a Short-Range Wnt Gradient in the Intestinal Stem-Cell Niche. *Nature*, **2016**, *530*, 340–

343.

[37] Nallet-Staub, F.; Yin, X.; Gilbert, C.; Marsaud, V.; Ben Mimoun, S.; Javelaud, D.; Leof, E.B.; Mauviel, A. Cell Density Sensing Alters TGF-B Signaling in a Cell-Type-Specific Manner, Independent From Hippo Pathway Activation. *Dev Cell*, **2015**, *32*, 640–651.

[38] Rosowski, K.A.; Mertz, A.F.; Norcross, S.; Dufresne, E.R.; Horsley, V. Edges of Human Embryonic Stem Cell Colonies Display Distinct Mechanical Properties and Differentiation Potential. *Sci Rep*, **2015**, *5*, 14218.

[39] Blin, G.; Picart, C.; Théry, M.; Puceat, M. Geometrical Confinement Guides Brachyury Self-Patterning in Embryonic Stem Cells. *bioRxiv*, **2017**, 1–46.

[40] Lee, G.Y.; Kenny, P.A.; Lee, E.H.; Bissell, M.J. Three-Dimensional Culture Models of Normal and Malignant Breast Epithelial Cells. *Nat Meth*, **2007**, *4*, 359–365.

[41] Lei, Y.; Schaffer, D.V. A Fully Defined and Scalable 3D Culture System for Human Pluripotent Stem Cell Expansion and Differentiation. *Proc Natl Acad Sci USA*, **2013**, *110*, E5039–E5048.

[42] Coucouvanis, E.; Martin, G.R. Signals for Death and Survival: a Two-Step Mechanism for Cavitation in the Vertebrate Embryo. *Cell*, **1995**, *83*, 279–287.

[43] Meinhardt, A.; Eberle, D.; Tazaki, A.; Ranga, A.; Niesche, M.; Wilsch-Bräuninger, M.; Stec, A.; Schackert, G.; Lutolf, M.; Tanaka, E.M. Stem Cell Reports. *Stem Cell Reports*, **2014**, *3*, 1–13.

[44] Ranga, A.; Girgin, M.; Meinhardt, A.; Eberle, D.; Caiazzo, M.; Tanaka, E.M.; Lutolf, M.P. Neural Tube Morphogenesis in Synthetic 3D Microenvironments. *Proc Natl Acad Sci USA*, **2016**, *113*, E6831–E6839.

[45] Engler, A.J.; Sen, S.; Sweeney, H.L.; Discher, D.E. Matrix Elasticity Directs Stem Cell Lineage Specification. *Cell*, **2006**, *126*, 677–689.

[46] Przybyla, L.; Lakins, J.N.; Weaver, V.M. Tissue Mechanics Orchestrate Wnt-Dependent Human Embryonic Stem Cell Differentiation. *Cell Stem Cell*, **2016**, *19*, 462–475.

[47] Mendjan, S.; Mascetti, V.L.; Ortmann, D.; Ortiz, M.; Karjosukarso, D.W.; Ng, Y.; Moreau, T.; Pedersen, R.A. NANOG and CDX2 Pattern Distinct Subtypes of Human Mesoderm During Exit From Pluripotency. *Cell Stem Cell*, **2014**.

[48] Sun, Y.; Yong, K.M.A.; Villa-Diaz, L.G.; Zhang, X.; Chen, W.; Philson, R.; Weng, S.; Xu, H.; Krebsbach, P.H.; Fu, J. Hippo/YAP-Mediated Rigidity-Dependent Motor Neuron Differentiation of Human Pluripotent Stem Cells. *Nat Mater*, **2014**, *13*, 599–604.

[49] Li, Y.; Moretto-Zita, M.; Soncin, F.; Wakeland, A.; Wolfe, L.; Leon-Garcia, S.; Pandian, R.; Pizzo, D.; Cui, L.; Nazor, K.; Loring, J.F.; Crum, C.P.; Laurent, L.C.; Parast, M.M. BMP4-Directed Trophoblast Differentiation of Human Embryonic Stem Cells Is Mediated Through a ΔNp63+ Cytotrophoblast Stem Cell State. *Development*, **2013**, *140*, 3965–3976.

[50] Horii, M.; Li, Y.; Wakeland, A.K.; Pizzo, D.P.; Nelson, K.K.; Sabatini, K.; Laurent, L.C.; Liu, Y.; Parast, M.M. Human Pluripotent Stem Cells as a Model of Trophoblast Differentiation in Both Normal Development and Disease. *Proc Natl Acad Sci USA*, **2016**, 201604747.

[51] Amita, M.; Adachi, K.; Alexenko, A.P.; Sinha, S.; Schust, D.J.; Schulz, L.C.; Roberts, R.M.; Ezashi, T. Complete and Unidirectional Conversion of Human Embryonic Stem Cells to Trophoblast by BMP4. *Proc Natl Acad Sci USA*, **2013**, *110*, E1212–E1221.

[52] Yang, Y.; Adachi, K.; Sheridan, M.A.; Alexenko, A.P.; Schust, D.J.; Schulz, L.C.; Ezashi, T.; Roberts, R.M. Heightened Potency of Human Pluripotent Stem Cell Lines

[53] Nemashkalo, A.; Ruzo, A.; Heemskerk, I.; Warmflash, A. Morphogen and Community Effects Determine Cell Fates in Response to BMP4 Signaling in Human Embryonic Stem Cells. *Development*, **2017**, dev.153239.

Created by Transient BMP4 Exposure. *Proc Natl Acad Sci USA*, **2015**.

[54] Nakamura, T.; Mine, N.; Nakaguchi, E.; Mochizuki, A.; Yamamoto, M.; Yashiro, K.; Meno, C.; Hamada, H. Generation of Robust Left-Right Asymmetry in the Mouse Embryo Requires a Self-Enhancement and Lateral-Inhibition System. *Dev Cell*, **2006**, *11*, 495–504.

[55] Müller, P.; Rogers, K.W.; Jordan, B.M.; Lee, J.S.; Robson, D.; Ramanathan, S.; Schier, A.F. Differential Diffusivity of Nodal and Lefty Underlies a Reaction-Diffusion Patterning System. *Science*, **2012**, *336*, 721–724.

[56] Tewary, M.; Ostblom, J.E.; Shakiba, N.; Zandstra, P.W. A Defined Platform of Human Peri-Gastrulation-Like Biological Fate Patterning Reveals Coordination Between Reaction-Diffusion and Positional-Information. *bioRxiv*, **2017**, 1–41.

[57] Nemashkalo, A.; Ruzo, A.; Heemskerk, I.; Warmflash, A. Morphogen and Community Effects Determine Cell Fates in Response to BMP4 Signaling in Human Embryonic Stem Cells. *bioRxiv*, **2017**, 1–30.

[58] Gurdon, J.B. A Community Effect in Animal Development. *Nature*, **1988**, *336*, 772–774.

[59] Bolouri, H.; Davidson, E.H. The Gene Regulatory Network Basis of the "Community Effect," and Analysis of a Sea Urchin Embryo Example. *Dev Biol*, **2010**, *340*, 170–178.

[60] Saka, Y.; Lhoussaine, C.; Kuttler, C.; Ullner, E.; Thiel, M. Theoretical Basis of the Community Effect in Development. *BMC Syst Biol*, **2011**, *5*, 54.

[61] Berge, ten, D.; Koole, W.; Fuerer, C.; Fish, M.; Eroglu, E.; Nusse, R. Wnt Signaling Mediates Self-Organization and Axis Formation in Embryoid Bodies. *Cell Stem Cell*, **2008**, *3*, 508–518.

[62] Turner, D.; Alonso-Crisostomo, L.; Girgin, M.; Baillie-Johnson, P.; Glodowski, C.R.; Hayward, P.C.; Collignon, J.; Gustavsen, C.; Serup, P.; Steventon, B.; Lutolf, M.; Martinez Arias, A. Gastruloids Develop the Three Body Axes in the Absence of Extraembryonic Tissues and Spatially Localised Signalling. *bioRxiv*, **2017**, 1–26.

[63] Nowotschin, S.; Costello, I.; Piliszek, A.; Kwon, G.S.; Mao, C.-A.; Klein, W.H.; Robertson, E.J.; Hadjantonakis, A.-K. The T-Box Transcription Factor Eomesodermin Is Essential for AVE Induction in the Mouse Embryo. *Genes Dev*, **2013**, *27*, 997–1002.

[64] Migeotte, I.; Omelchenko, T.; Hall, A.; Anderson, K.V. Rac1-Dependent Collective Cell Migration Is Required for Specification of the Anterior-Posterior Body Axis of the Mouse. *PLoS Biol*, **2010**, *8*, e1000442.

[65] Gjorevski, N.; Ranga, A.; Lutolf, M.P. Bioengineering Approaches to Guide Stem Cell-Based Organogenesis. *Development*, **2014**, *141*, 1794–1804.

A. Bra/Sox2/Sox1::GFP 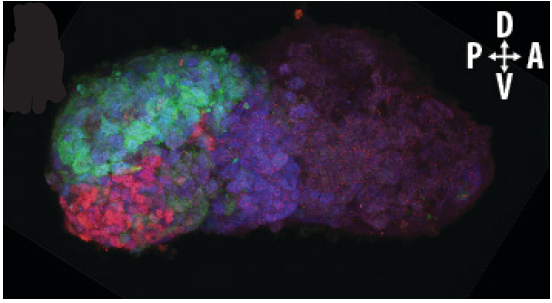 B. E-Cad/Hoechst/β-cat 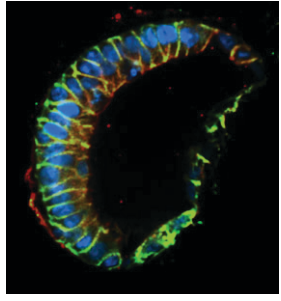 C. DAPI/OCT4/Stella GFP 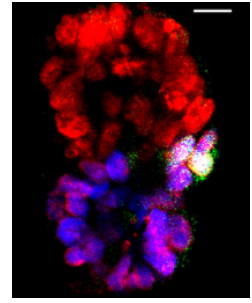